\documentclass[trackchanges,twocolumn]{aastex7}

\usepackage{CJK}
\usepackage{framed,amsmath,bm,booktabs}
\usepackage{float}

\graphicspath{{./}{figures}}

\submitjournal{ApJ}
\shorttitle{FZH Last Crossing}
\shortauthors{Hanjue Zhu}

\begin{document}
\begin{CJK*}{UTF8}{gkai}

\title{The Last Crossing in Excursion-Set Theory of Cosmic Reionization}

\correspondingauthor{Hanjue Zhu (朱涵珏)}
\author[orcid=0000-0003-0861-0922,sname='Zhu']{Hanjue Zhu (朱涵珏)}
\affiliation{Department of Astronomy \& Astrophysics; 
The University of Chicago; 
Chicago, IL 60637, USA}
\email[show]{hanjuezhu@uchicago.edu}

\begin{abstract}
I introduce the last crossing of the photon-counting barrier as a statistic for
analytical models of cosmic reionization. Because ionizing galaxies are biased
tracers of density and their photons propagate through gas of spatially varying
opacity, the galaxy--IGM connection requires statistics beyond the global
ionized fraction. The excursion-set model of \citet{Furlanetto2004}
identifies the ionized-bubble scale around a point with the first crossing of a
photon-counting barrier. I develop the last-crossing formalism and define the
last-crossing scale, $R_\ell$, as the smallest scale on which the enclosed
photon budget still ionizes the enclosed gas. I derive its distribution
analytically. Using the empirical-barrier framework of \citet{Kaurov2016}
applied to CROC simulations, I measure both crossings in the simulated IGM.
The last-crossing formalism separates externally ionized regions, whose
photon-counting condition fails below $R_\ell$, from internally ionized regions
whose trajectories remain above the barrier to the resolution scale. Resolved
external regions are predominantly underdense, with larger $R_\ell$ selecting
lower density and later reionization. At fixed local density, first-crossing
scale, and reionization redshift, these regions show deficits in
ionizing-luminosity and galaxy-number density at small radius relative to
matched unresolved regions, and excesses in both at intermediate radius. Their
combination of low gas density and a deficit of nearby sources suggests that
Ly$\alpha$ transmission spikes may preferentially arise in these regions: low
density reduces the opacity, while sources on larger scales maintain the
ionized state. The last-crossing picture connects the internal source geometry
of ionized regions to a testable prediction for radial galaxy distributions
around transmission-selected IGM locations.
\end{abstract}

\section{Introduction}
 
Cosmic reionization is the cosmological phase transition during which the
intergalactic medium (IGM), neutral after recombination, was reionized by
ultraviolet photons from the first generations of stars, galaxies, and quasars
\citep[for reviews, see][]{Barkana2001,Furlanetto2006,McQuinn2016,
GnedinMadau2022}. Its integrated history is constrained by the Thomson optical depth measured from the cosmic microwave background \citep{Planck2020}, while the disappearance of long Gunn--Peterson troughs and the emergence of widespread Ly$\alpha$ transmission constrain the late stages of reionization \citep{Becker2015,Bosman2022}. Many theoretical and semi-numerical descriptions
therefore use the volume-averaged ionized fraction
$\langle x_{\rm HII}\rangle(z)$ as the basic coordinate for comparing models.
That coordinate leaves out where ionized regions sit, how galaxies are arranged
relative to the gas they ionize, how far ionizing photons propagate before
being absorbed, and how strongly the ionization field remembers the underlying
density field. Two models with identical $\langle x_{\rm HII}\rangle(z)$ can
produce substantially different radiation fields, mean free paths, and
observable structures.

Reionization is a radiative-transfer problem in an inhomogeneous medium.
Galaxies, the dominant sources of ionizing photons during the epoch
\citep{Robertson2015,Finkelstein2019}, form preferentially in overdense regions
and are biased tracers of the underlying density field \citep{Mo1996}. The gas
through which their photons propagate is structured too: dense in the
filaments and halos that host the sources, more rapidly recombining in the
densest environments, and more transparent in the low-density channels between
them. The morphology of ionized regions, the patchiness of the ionizing
background, and the spatial relation between galaxies and ionized gas depend on
how sources, sinks, and large-scale structure are arranged across scale, not
just on the total number of photons produced.

Current and upcoming observations are beginning to probe this scale dependence
directly. The redshifted 21\,cm signal targets fluctuations in neutral hydrogen
and aims to measure the large-scale topology of ionized regions
\citep{Pritchard2012,Mellema2013}, and present experiments already place upper
limits on its power spectrum during reionization
\citep{Mertens2020,Abdurashidova2023}. The patchy kinetic
Sunyaev--Zel'dovich signal responds to the spatial distribution and bulk motion
of ionized gas \citep{McQuinn2005,Mesinger2012}. Ly$\alpha$ damping wings, dark
gaps, and transmission spikes probe the density, ionization, radiation, and
thermal fields along individual quasar sightlines
\citep{Becker2015,Bosman2022}. JWST programs in quasar fields, including
EIGER and ASPIRE, now connect those sightlines to the surrounding galaxy
distribution \citep{Kashino2023,Wang2023,Jin2024,Kakiichi2025}.
Galaxy--IGM cross-correlations and forward models are therefore becoming a
direct test of whether transmission is controlled by nearby galaxies, by the
larger ionized environment, or by both
\citep{Kakiichi2018,Meyer2020,Conaboy2025,Garaldi2025Observability,
Garaldi2025Opacity}.

Radiative-transfer simulations evolve sources, gas, radiation, and
recombinations together, producing three-dimensional ionization fields that
can be compared directly with observations
\citep[e.g.,][]{Gnedin2014,Ocvirk2016,Rosdahl2018,Kannan2022,
Garaldi2022,Smith2022,Garaldi2023,Zier2026}. These fields show where ionized
and neutral regions lie; identifying which physical ingredients set their
sizes, shapes, and relation to the galaxy distribution requires more
controlled approaches. Analytic and semi-analytic models supply that control by
introducing simplified treatments of source formation, recombinations,
radiative transfer, and topology, isolating which physical ingredients are
assumed to set the ionization morphology
\citep{ChoudhuryFerrara2005,Mesinger2011,Zahn2011,SobacchiMesinger2014}.

The excursion-set model of \citet{Furlanetto2004} (hereafter FZH04) is the
canonical example of this approach applied to ionized-bubble morphology. FZH04
connects ionized regions to large-scale density fluctuations: overdense regions
host more collapsed halos and therefore more ionizing sources. Excursion-set
theory turns this picture into a trajectory problem. Around each point, the
density field is smoothed on a sequence of radii, and a region is identified as
ionized when the sources enclosed by that smoothing aperture can ionize the gas
enclosed by the same aperture. FZH04 thereby reduces the growth of ionized
structure to a barrier-crossing calculation, yielding predictions for the
bubble size distribution, the ionized fraction, and the large-scale 21\,cm
power that have served as a touchstone for both simulations and semi-numerical
models
\citep{McQuinn2007,Mesinger2007,Mesinger2011,Lin2016,GnedinMadau2022}.

The standard use of this barrier is the \emph{first} crossing. Following a trajectory from large to small smoothing radii, the first scale satisfying the photon-counting condition gives the largest self-ionizing region around the point---the excursion-set bubble scale that connects the biased galaxy distribution to the large H~{\sc ii} regions targeted by 21\,cm measurements. The same trajectory, however, contains more information. Once the barrier has been crossed, the trajectory can be followed inward. The final crossing encountered as the smoothing radius decreases defines the resolved \emph{last} crossing, while trajectories that remain above the barrier at the minimum smoothing scale belong to the finite-endpoint population. The last crossing measures how far inward the local source-counting condition continues to hold, and partitions ionized regions by the scale on which the FZH04 enclosed-source photon budget finally stops closing.

In this paper, I develop the last-crossing statistic for the FZH04 barrier,
measure it in a radiative-transfer simulation, and examine the physical
environments the statistic selects.
Section~\ref{sec:last_crossing_FZH04} sets up the finite-endpoint last-crossing
calculation. Section~\ref{sec:theory_results} presents the analytic first- and
last-crossing distributions and their evolution through reionization.
Section~\ref{sec:simulation_barriers} extracts an empirical barrier from CROC
trajectories following \citet{Kaurov2016} and compares simulated and analytic
crossing distributions. Section~\ref{sec:resolved_Rell_physics} examines the
environments selected by resolved versus unresolved last crossings.
Section~\ref{sec:observational_program} measures their source environments and
draws out the implications for the galaxy--IGM connection.
Section~\ref{sec:summary} concludes.
Appendix~\ref{app:mc_validation} validates the deterministic last-crossing
solver against direct random-walk realizations.

\section{Last Crossing of the FZH04 Barrier}
\label{sec:last_crossing_FZH04}
 
The last-crossing calculation rests on the same photon-counting condition as the original FZH04 model; only the crossing rule changes, from the largest smoothing radius satisfying the condition to the smallest. What follows is a self-contained derivation of the last-crossing equations, drawing on the relevant elements of structure formation theory and FZH04.
 
Recall the definitions from the excursion-set formalism. Let $\delta(\mathbf{x})$ be the initial linear density contrast, modeled as a statistically homogeneous Gaussian random field. Its smoothing on comoving radius $R$ is
\begin{equation}
    \delta_R(\mathbf{x})
    =
    \int d^3x'\,
    W_R(\mathbf{x}-\mathbf{x}')\,\delta(\mathbf{x}') ,
\end{equation}
with $W_R$ a window normalized to unit volume. In Fourier space, the convolution becomes a multiplication, $\widetilde{\delta_R}(\mathbf{k})=\widetilde W(kR)\widetilde\delta(\mathbf{k})$, and the variance of the smoothed field follows directly from the definition of the linear matter power spectrum, $\langle\widetilde\delta(\mathbf{k})\widetilde\delta^*(\mathbf{k}')\rangle=(2\pi)^3\delta_D(\mathbf{k}-\mathbf{k}')P(k)$:
\begin{equation}
    S(R)
    \equiv
    \left\langle \delta_R^2 \right\rangle
    =
    \int_0^\infty \frac{k^2\,dk}{2\pi^2}\,
    P(k)\,|\widetilde W(kR)|^2 .
\end{equation}
$S(R)$ measures the fluctuation power retained after smoothing on scale $R$. It decreases monotonically with $R$: large radii average over many independent modes and yield small $S$, while small radii retain more power and yield large $S$. At a fixed position, the sequence $\delta_R(\mathbf{x})$ as $R$ decreases defines a trajectory $\delta(S)$ in the variance coordinate.
 
The FZH04 model estimates the abundance of ionizing sources from the conditional collapsed fraction. In extended Press--Schechter theory \citep{BondEtal1991}, a region collapses into a halo of mass $M$ when its smoothed density on the corresponding scale first crosses the linear threshold $\delta_c(z)$. For a Gaussian field smoothed with a sharp-$k$ window, the random walk $\delta(S)$ is Markovian, and the first-crossing distribution above a constant threshold follows from the absorbing-barrier construction of \citet{BondEtal1991}: once a trajectory reaches $\delta_c(z)$, the corresponding mass is counted as collapsed and removed from the unbound population, so that no mass element is counted twice on smaller smoothing scales. Conditioning on a parent region of overdensity $\delta_R$ at variance $S$ gives the fraction of mass collapsed into halos above a minimum source mass $M_{\min}$:
\begin{equation}
    f_{\rm coll}(\delta_R,S;z)
    =
    \mathrm{erfc}
    \left[
    \frac{\delta_c(z)-\delta_R}
    {\sqrt{2(S_{\min}-S)}}
    \right],
\end{equation}
where $\delta_c(z)$ is the linear collapse threshold and $S_{\min}\equiv S(M_{\min})$ is the variance at the minimum source-halo scale. The quantity $S_{\min}-S$ is the variance contributed by density modes between the smoothed region and the halo scale; it controls how much sub-smoothing-scale fluctuation power can boost the local collapsed fraction.
 
The ionizing efficiency $\zeta$, which encodes the number of ionizing photons produced per baryon of collapsed mass, then enters through the photon-counting condition: the smoothed region is ionized when its source population can supply enough photons to ionize the enclosed gas,
\begin{equation}
    \zeta\,f_{\rm coll}(\delta_R,S;z)\ge 1 .
\end{equation}
Solving for the overdensity threshold gives
\begin{equation}
    \delta_R
    \ge
    \delta_c(z)
    -
    \sqrt{2}\,K(\zeta)\sqrt{S_{\min}-S},
\end{equation}
where $K(\zeta)\equiv \mathrm{erfc}^{-1}(\zeta^{-1})$. For $\zeta>1$, $K(\zeta)>0$, so the right-hand side lies below $\delta_c$ for $S<S_{\min}$ and rises to $\delta_c$ at the endpoint $S_{\min}$. This defines the FZH04 barrier,
\begin{equation}
    B(S,z)
    =
    \delta_c(z)
    -
    \sqrt{2}\,K(\zeta)\sqrt{S_{\min}-S},
\end{equation}
and a smoothed region satisfies the FZH04 ionization condition when $\delta(S)\ge B(S,z)$.
 
The first crossing is the smallest variance, equivalently the largest radius, at which the trajectory satisfies this condition,
\begin{equation}
    S_f
    =
    \min\{S:\delta(S)\ge B(S,z)\},
\end{equation}
with associated radius $R_f=R(S_f)$, the largest self-ionizing region around the point. The last crossing is its conjugate,
\begin{equation}
    S_\ell
    =
    \max\{S:\delta(S)\ge B(S,z)\},
\end{equation}
with $R_\ell=R(S_\ell)$ the smallest smoothing scale on which the FZH04 condition is still met.
 
\begin{figure*}
    \centering
    \includegraphics[width=\textwidth]{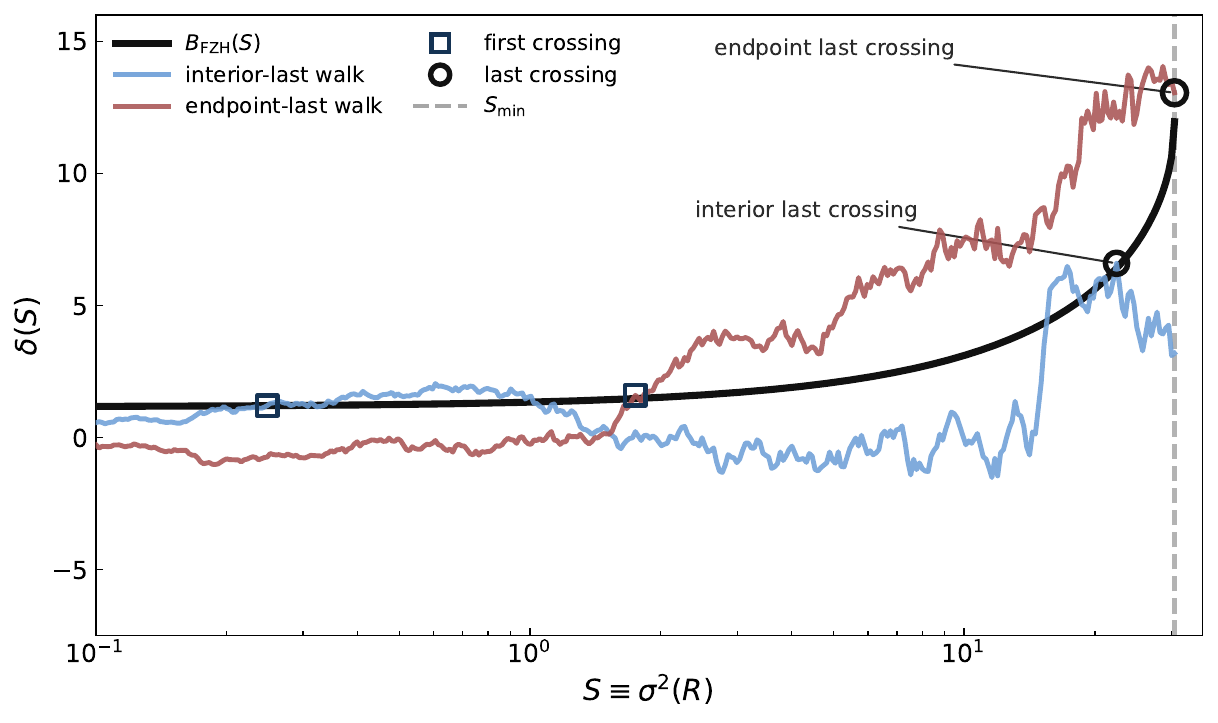}
    \caption{
    Example sharp-$k$ trajectories crossing the FZH04 barrier in $S$-space.
    The horizontal axis is $S \equiv \sigma^2(R)$, so moving to the right
    corresponds to decreasing smoothing radius. The black curve is the FZH04
    barrier, and the vertical dashed line marks the endpoint $S_{\min}$.
    Squares mark first crossings; circles mark last crossings. The blue
    trajectory drops below the barrier at scales smaller than its last
    crossing and ends below it at $S_{\min}$, giving an interior last crossing.
    The red trajectory remains above the barrier at $S_{\min}$ and therefore
    contributes to the endpoint probability.
    }
    \label{fig:walk_example}
\end{figure*}

Figure~\ref{fig:walk_example} illustrates these definitions. The first crossing is the smallest-$S$ point at which the trajectory enters the ionized side of the barrier; the last crossing is the largest-$S$ point at which it still lies on that side. Trajectories whose last crossing falls before $S_{\min}$ (the blue example) contribute a smooth interior distribution of $S_\ell$. Those that remain above the barrier at $S_{\min}$ (the red example) populate a discrete endpoint contribution at $R_{\min}$. This split between a smooth interior density and a discrete endpoint atom is specific to the FZH04 barrier, where the source model fixes a finite small-scale endpoint.
 
Figure~\ref{fig:walk_example} makes clear that the last-crossing scale is not determined by the trajectory's value at a single $S$. A trajectory can touch the barrier at some $S'$ without that being its last crossing, if it crosses again at a larger variance. Identifying the last crossing therefore requires the joint behavior of $\delta(S)$ across an interval of smoothing scales, not the one-point probability at $S'$ alone. Concretely, for any coarser $S<S'$ the calculation needs the probability that a trajectory passing through the barrier at $S'$ was already above it at $S$ --- a question about the joint statistics of $\delta(S)$ and $\delta(S')$.
 
A sharp-$k$ filter on the density field, $\widetilde W(kR) = 1$ for $k<1/R$ and $0$ otherwise, makes those statistics simple. Each step in $S$ adds Fourier modes in a narrow shell that are uncorrelated with the modes at smaller $S$, so $\delta(S)$ is a Markov Gaussian walk. The pair $(\delta(S),\delta(S'))$ with $S<S'$ is therefore jointly Gaussian with
\begin{equation}
    \langle \delta(S)^2\rangle=S,
    \qquad
    \langle \delta(S')^2\rangle=S',
    \qquad
    \langle \delta(S)\delta(S')\rangle=S .
\end{equation}
The result needed below is the conditional distribution: given $\delta(S')=\delta'$ at $S'>S$, the value of $\delta(S)$ is Gaussian with mean $(S/S')\delta'$ and variance $S(S'-S)/S'$.
 
\subsection{The Hopkins last-crossing equation}
\label{sec:hopkins_review}
 
\citet{Hopkins2012} derived a Volterra integral equation for the last-crossing distribution of sharp-$k$ trajectories in the context of the stellar initial mass function. The original problem uses a lognormal turbulent density field, with $\delta(S)$ a smoothed logarithmic density contrast and $B(S)$ the critical density for self-gravity; the first and last crossings then correspond respectively to the parent self-gravitating cloud and to the smallest self-gravitating fragments within it. The physical field and barrier therefore differ from the reionization problem, but the sharp-$k$ crossing mathematics is the same. I sketch the derivation here to make explicit which parts transfer to our setting and which require modification.
 
With no barrier in the problem, the trajectory at variance $S$ has the Gaussian density
\begin{equation}
    P_0(\delta|S)=\frac{1}{\sqrt{2\pi S}}\exp\!\left(-\frac{\delta^2}{2S}\right),
\end{equation}
and the calculation proceeds inward: trajectories start from a small initial radius $R_i$, corresponding to large $S_i$, and the variance grid is walked toward smaller $S$ (larger physical scales). The Hopkins setup places $S_i$ in a regime where
\begin{equation}
    P[\delta(S_i)>B(S_i)]\to 0,
    \label{eq:hopkins_endpoint}
\end{equation}
which serves as the boundary condition for the derivation. Trajectories that cross the barrier then have their last crossing at some interior value $S<S_i$, and the last-crossing probability is described by a smooth density $f_\ell(S)$.
 
Let $\Pi(\delta|S)\,d\delta$ denote the probability that the trajectory has value between $\delta$ and $\delta+d\delta$ at variance $S$, \emph{conditional on not having crossed the barrier at any larger variance $S'>S$} along the inward walk from $S_i$. Probability conservation requires
\begin{equation}
    1=\int_S^{S_i}dS'\,f_\ell(S')
    +\int_{-\infty}^{B(S)}d\delta\,\Pi(\delta|S),
    \label{eq:hopkins_conservation}
\end{equation}
where the first term collects trajectories whose last crossing already occurred at some $S'>S$ (integrated over the possible crossing location), and the second collects trajectories still below the barrier at $S$ (integrated over their current value of $\delta$). To obtain a closed equation for $f_\ell$, $\Pi(\delta|S)$ itself must be expressed in terms of $f_\ell$. The Gaussian density $P_0(\delta|S)$ counts \emph{all} trajectories at $(\delta,S)$, including those that have already crossed the barrier at some $S'$ between $S$ and $S_i$ and then diffused back to $\delta$. Subtracting that population leaves trajectories that have not yet crossed,
\begin{equation}
    \Pi(\delta|S)
    =P_0(\delta|S)
    -\int_S^{S_i}dS'\,f_\ell(S')\,P_{01}\!\left[\delta(S)\,\big|\,B(S')\right],
    \label{eq:hopkins_pi}
\end{equation}
with $P_{01}[\delta(S)|\delta(S')]$ the conditional Gaussian density at $S$ given $\delta(S')$. The integral is the Gaussian probability of arriving at $\delta(S)$ from the barrier location $B(S')$, weighted by the density of crossings $f_\ell(S')$. Differentiating Equation~\eqref{eq:hopkins_conservation} with respect to $S$ and substituting Equation~\eqref{eq:hopkins_pi} yields the Hopkins Volterra equation,
\begin{equation}
    f_\ell(S)=g_1(S)+\int_S^{S_i}dS'\,f_\ell(S')\,g_2(S,S'),
    \label{eq:hopkins_volterra}
\end{equation}
with kernels
\begin{align}
    g_1(S)&=\left[2\frac{dB}{dS}-\frac{B(S)}{S}\right]P_0(B(S)|S),\\
    g_2(S,S')&=\left[\frac{B(S)-B(S')}{S-S'}+\frac{B(S)}{S}-2\frac{dB}{dS}\right]P_{01}[B(S)|B(S')].
\end{align}
This is a Volterra equation of the second kind: the unknown $f_\ell$ at $S$ depends only on $f_\ell$ at larger $S'$, so the equation can be solved by stepping outward from the endpoint. For a linear barrier $B(S)=B_0+\beta S$, the Hopkins solution is
\begin{equation}
    f_\ell(S)=\beta\,P_0(B(S)|S),
    \label{eq:hopkins_linear_solution}
\end{equation}
which vanishes for a constant barrier ($\beta=0$): a Brownian walk recrosses any constant level arbitrarily close to the endpoint, leaving no finite last-crossing scale.
 
\subsection{Adapting the construction to the FZH04 barrier}
\label{sec:fzh_finite_endpoint}
 
Two features of the FZH04 problem prevent direct application of the Hopkins equations. First, the trajectory cannot be continued beyond the minimum source-halo scale,
\begin{equation}
    S_i\equiv S_{\min},
\end{equation}
so the endpoint $R_i$ is at a finite, physically motivated value rather than at a small radius chosen so that Equation~\eqref{eq:hopkins_endpoint} holds. Second, the barrier at this endpoint is finite,
\begin{equation}
    B_i\equiv B(S_i,z)=\delta_c(z),
\end{equation}
so a nonzero fraction of trajectories sits above the barrier at $S_i$ and never recrosses on the way out. These trajectories are assigned $S_\ell=S_i$ and contribute a discrete endpoint probability
\begin{equation}
    p_{\rm end}
    =
    P[\delta(S_i)\ge B_i]
    =
    1-\Phi\!\left(\frac{B_i}{\sqrt{S_i}}\right),
\end{equation}
with $\Phi$ the standard normal CDF. The full last-crossing distribution is therefore the sum of a smooth interior density and a Dirac contribution at $R_{\min}$. The Hopkins derivation produces only the smooth part; I now construct the interior equation in a form that admits the finite-endpoint correction.
 
The smooth interior density is the contribution from trajectories that end below the barrier at $S_i$ but were above it at some larger physical scale. I now isolate this population in two steps. The first step expresses its probability as a two-dimensional Gaussian integral that can be evaluated directly. The second step rewrites the same probability as an integral over the last-crossing location, which is what we ultimately want to recover.
 
For any $S<S_i$, define
\begin{equation}
    A(S)
    \equiv
    P\!\left[\delta(S)\ge B(S),\,\delta(S_i)<B_i\right] ,
\end{equation}
the joint probability that the trajectory is \emph{above} the barrier at the larger physical scale $S$ and \emph{below} it at the endpoint $S_i$. In the $(S,\delta)$ plane, this corresponds to trajectories that pass through the upper region at $S$ and end in the lower region at $S_i$. Such a trajectory must cross the barrier at least once between $S$ and $S_i$, and the final such crossing is its last crossing.
 
Computing $A(S)$ directly is straightforward because it depends only on the two-time joint distribution of $\delta(S)$ and $\delta(S_i)$, which is bivariate Gaussian. Using the Markov walk covariances $\langle \delta(S)^2\rangle=S$, $\langle \delta(S_i)^2\rangle=S_i$, and $\langle \delta(S)\delta(S_i)\rangle=S$, the joint distribution has covariance matrix
\begin{equation}
    \mathbf{C}
    =
    \begin{pmatrix}
        S & S\\
        S & S_i
    \end{pmatrix},
\end{equation}
and $A(S)$ is the integral of the corresponding two-dimensional Gaussian over the region $\delta(S)\ge B(S)$, $\delta(S_i)<B_i$. The complementary population, $\delta(S_i)\ge B_i$, is absent in the Hopkins boundary condition Equation~\eqref{eq:hopkins_endpoint} but is finite here; it forms the endpoint atom $p_{\rm end}$. The condition $\delta(S_i)<B_i$ therefore isolates the smooth interior part of the FZH04 last-crossing distribution.
 
The second step rewrites this same probability by partitioning trajectories according to where their last crossing occurs. Let $f_\ell(S')\,dS'$ denote the probability that the last crossing lies between $S'$ and $S'+dS'$. A trajectory contributing to $A(S)$ is one that has its last crossing at \emph{some} $S'\in[S,S_i]$ and that was already above the barrier at the earlier scale $S$. To turn this into an equation, two ingredients are needed: the probability that the last crossing happens near $S'$, which is $f_\ell(S')\,dS'$, and the conditional probability that the same trajectory was above the barrier at $S$ given that it touched the barrier at $S'$.
 
The conditional probability comes from Gaussian conditioning. Fixing $\delta(S')=B(S')$ for $S'>S$, standard formulas for the conditional distribution of a jointly Gaussian pair give
\begin{equation}
    \langle \delta(S)\,|\,\delta(S')=B(S')\rangle = \frac{S}{S'}B(S'),
\end{equation}
\begin{equation}
    \mathrm{Var}[\delta(S)\,|\,\delta(S')=B(S')] = S - \frac{S^2}{S'} = \frac{S(S'-S)}{S'}.
\end{equation}
The conditional probability of being above the barrier at $S$ is then the upper tail of this Gaussian,
\begin{align}
    K(S,S')
    &=
    P\!\left[\delta(S)\ge B(S)\,\big|\,\delta(S')=B(S')\right] \\ \nonumber
    &=
    1-
    \Phi\!\left[
    \frac{B(S)-B(S')\,S/S'}
    {\sqrt{S(S'-S)/S'}}
    \right].
\end{align}
The product $f_\ell(S')\,dS'\,K(S,S')$ is therefore the probability that the last crossing occurs near $S'$ \emph{and} the trajectory was above the barrier at $S$. Summing over all possible last-crossing locations gives a Volterra integral equation for the interior last-crossing density,
\begin{equation}
    A(S)
    =
    \int_S^{S_i} dS'\,
    f_\ell(S')\,K(S,S') .
    \label{eq:fzh_volterra}
\end{equation}
 
The Markov property closes this equation. Because the sharp-$k$ walk is Markovian, the part of the trajectory at $S<S'$ is independent of the later segment $S'>S''>S_i$ once the value at $S'$ is fixed. The requirement that no additional crossing occurs between $S'$ and $S_i$ therefore changes the weight assigned to $S'$ --- this is precisely what $f_\ell(S')$ encodes --- but it does not change the conditional distribution of $\delta(S)$ at coarser scales. The kernel $K(S,S')$, conditioned only on $\delta(S')=B(S')$, captures the full statistics of the earlier segment.
 
Equation~\eqref{eq:fzh_volterra} differs from the Hopkins Volterra equation, Equation~\eqref{eq:hopkins_volterra}, in two ways. First, it is a Volterra equation of the first kind: $f_\ell$ appears only inside the integral, in contrast to Hopkins's second-kind form where $f_\ell$ appears on both sides. The left-hand side $A(S)$ is no longer the unknown itself but a known function, the joint probability that the trajectory is above the barrier at $S$ \emph{and} below it at $S_i$, computable directly from the bivariate Gaussian. The endpoint condition is built into $A(S)$ rather than imposed through a boundary limit. Second, the upper limit of integration is the physical scale $S_i=S_{\min}$, rather than a small-radius cutoff chosen so that no trajectory remains above the barrier at $S_i$. The kernel $K(S,S')$ is the standard sharp-$k$ conditional probability and is the same as in the Hopkins setup. The interior density $f_\ell$ is then recovered by inverting Equation~\eqref{eq:fzh_volterra} numerically.

\subsection{Numerical solution}
\label{sec:numerical}
 
I solve Equation~\eqref{eq:fzh_volterra} by binning the possible last-crossing locations from the endpoint outward. After discretization the equation becomes a triangular linear system: $A_n$ at the outer edge of bin $n$ depends only on the unknowns $p_m$ with $m\le n$, so the system can be inverted by forward substitution without storing a full matrix. Let $p_n$ be the probability that the interior last crossing falls in bin $n$, and $A_n$ the value of $A(S)$ at the outer edge of bin $n$. Starting from the innermost bin,
\begin{equation}
    A_1=p_1K_{11},
    \qquad
    p_1=\frac{A_1}{K_{11}},
\end{equation}
adding the next bin,
\begin{equation}
    A_2=p_1K_{21}+p_2K_{22},
    \qquad
    p_2=\frac{A_2-p_1K_{21}}{K_{22}},
\end{equation}
and continuing by the general recursion,
\begin{equation}
    p_n
    =
    \frac{A_n-\sum_{m=1}^{n-1}p_mK_{nm}}{K_{nn}} ,
\end{equation}
where $K_{nm}$ is the kernel evaluated at the outer edge of bin $n$ and at the midpoint of bin $m$. At each step, the previously solved bins account for part of $A_n$, and the remainder is assigned to the new bin.
 
Casting the grid in $u\equiv\sqrt{S_i-S}$ rather than directly in $S$ linearizes the FZH04 barrier, $B(u)=\delta_c(z)-\sqrt{2}\,K(\zeta)\,u$, so that uniform spacing in $u$ resolves the rapid variation of $B(S,z)$ near $S_{\min}$, where $dB/dS$ diverges. The endpoint sits at $u=0$, and successive bins march outward to larger radii.
 
The total probability decomposes as
\begin{equation}
    p_{\rm none}+q_{\rm int}+p_{\rm end}=1,
\end{equation}
where $p_{\rm none}$ is the probability that the trajectory never crosses the barrier, $p_{\rm end}$ is the endpoint contribution, and $q_{\rm int}=\sum_n p_n$ is the interior crossing probability. By construction, $q_{\rm int}+p_{\rm end}$ equals the probability that the trajectory satisfies the FZH04 condition at least once, which equals the first-crossing probability and serves as a normalization check on the recursion. The bin probabilities give a density per logarithmic radius,
\begin{equation}
    \frac{dP}{d\ln R}
    \simeq
    \frac{p_n}{|\ln(R_{\rm right}/R_{\rm left})|},
\end{equation}
where $R_{\rm left}$ and $R_{\rm right}$ bound the bin. The endpoint probability $p_{\rm end}$ is reported separately at $R_{\min}$ rather than folded into this smooth density.

\section{Analytic Crossing Distributions}
\label{sec:theory_results}

For any choice of redshift, ionizing efficiency, cosmology, and minimum
source mass, the Volterra solver yields both the first- and last-crossing
scale distributions. The first-crossing distribution describes the
enclosing ionized scale around a random point; the last-crossing
distribution describes the smallest scale on which the same trajectory
still satisfies the photon-counting condition. Together they probe the
full source-counting hierarchy of a single ionized region.

Both distributions are normalized to the ionized volume fraction predicted
by the analytic FZH04 model rather than to unity. Denoting this fraction by
$\langle x_{\rm HII}\rangle_{\rm th}$, the first-crossing density integrates
as
\begin{equation}
    \int d\ln R\, \frac{dP_f}{d\ln R}
    = q_{\rm first}
    \equiv \langle x_{\rm HII}\rangle_{\rm th},
\end{equation}
where $q_{\rm first}$ is the probability that a random point belongs to
at least one self-ionizing FZH04 region. The interior last-crossing density
and the endpoint contribution together integrate to the same total,
\begin{equation}
    \int d\ln R\, \frac{dP_{\ell,{\rm int}}}{d\ln R}
    + p_{\rm end}
    = q_{\rm first}.
\end{equation}

\begin{figure}[htb!]
    \centering
    \includegraphics[width=\linewidth]{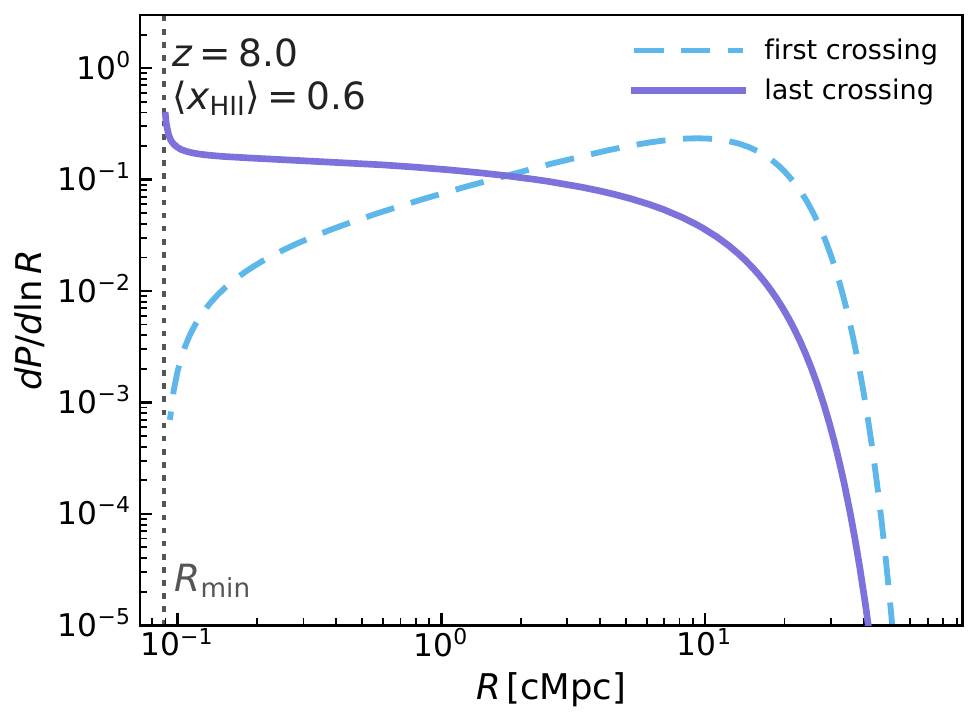}
    \caption{
    First- and last-crossing length distributions for a representative
    FZH04 model at $z=8.0$ and $\langle x_{\rm HII}\rangle=0.6$. Both
    curves are computed from the same barrier. The first crossing gives
    the largest smoothing radius at which the FZH04 photon-counting
    condition $\zeta f_{\rm coll}\ge 1$ is satisfied; the last crossing
    gives the smallest such radius. Both curves integrate to the
    theoretical ionized volume fraction rather than to unity.
    }
    \label{fig:crossing_length_distributions}
\end{figure}

Figure~\ref{fig:crossing_length_distributions} shows the resulting
probability densities. Both curves come from the same barrier and the
same ensemble of theory trajectories; only the crossing rule changes.
The first-crossing distribution peaks at larger radii because it selects
the largest self-ionizing region around each point, while the
last-crossing distribution peaks at smaller radii because it follows the
same condition inward to the scale at which it last holds.
Appendix~\ref{app:mc_validation} validates the deterministic
last-crossing calculation against direct Monte Carlo realizations of the
same sharp-$k$ process.

\begin{figure*}
    \centering
    \includegraphics[width=\textwidth]{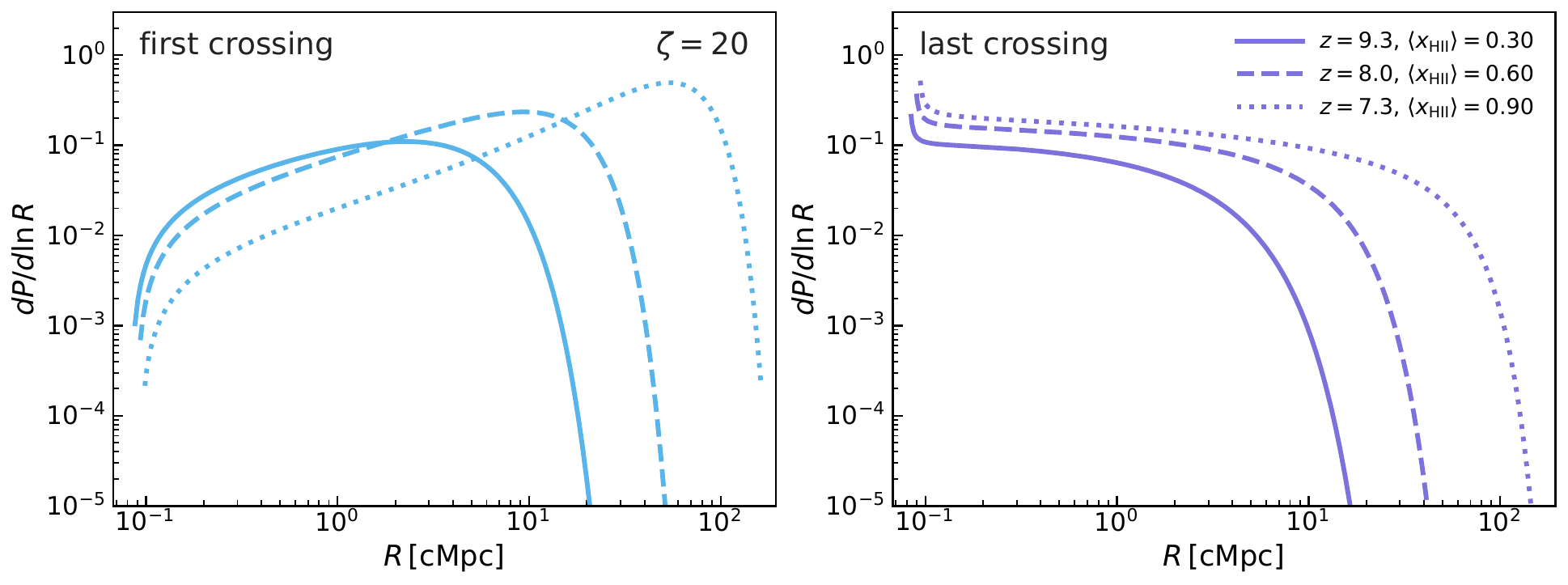}
    \caption{
    Evolution of the first- and last-crossing length distributions in the
    analytic FZH04 model at different stages of cosmic reionization. The
    first-crossing distribution shifts to larger radii as the ionized
    fraction grows, mirroring the expansion of enclosing ionized regions.
    The last-crossing distribution evolves in tandem: its large-scale
    extent also moves outward, while its probability remains concentrated
    at smaller radii. The coordinated evolution reflects an approximately
    self-similar source-counting hierarchy.
    }
    \label{fig:length_evolution}
\end{figure*}

Figure~\ref{fig:length_evolution} tracks the evolution of the two
distributions at different stages of reionization. As the ionized
fraction grows, the first-crossing distribution moves to larger radii,
mirroring the expansion of enclosing ionized regions predicted by FZH04.
The last-crossing distribution evolves in an approximately self-similar
manner: its large-scale extent moves outward together with the
characteristic first-crossing scale, while its probability remains
weighted toward smaller radii. This behavior reflects the approximately
scale-free hierarchy of the initial density field. Over the range where
the matter power spectrum is locally close to a power law, rescaling the
smoothing radius changes the amplitude of the fluctuations without
introducing a preferred physical length, allowing the crossing
distributions to expand while approximately preserving their relative
form. Exact self-similarity is broken by the finite endpoint at
$S_{\min}$ and by the scale-dependent slope of the matter power spectrum.

\section{Empirical Barrier in a Radiative-Transfer Simulation}
\label{sec:simulation_barriers}

The analytic FZH04 barrier cannot be applied directly to a
radiative-transfer simulation. The simulated ionization field depends on
source discreteness, recombinations, photon propagation, and the nonlocal
radiation field, none of which enter the analytic photon-counting condition.
Following \citet{Kaurov2016}, I therefore construct an empirical barrier from
the simulation itself and apply the same crossing analysis to it.

I construct the trajectories from the binned, log-transformed, and
Gaussianized density field using a sharp-$k$ Fourier filter on logarithmically
spaced radii from $10^{-3}L_{\rm box}$ to $L_{\rm box}$. At each radius, I
divide grid locations into ionized and neutral samples according to whether
their reionization scale factor lies before or after the snapshot epoch. The
class-conditional trajectory distributions are weighted by the corresponding
volume fractions. The empirical barrier is the trajectory value at which the
prior-weighted neutral and ionized distributions are equal, or equivalently at
which a location with that trajectory value is equally likely to belong to
either class. Using the full simulation volume yields a well-sampled estimate
of this barrier at each smoothing radius.

The simulated barrier is fuzzy. In the analytic FZH04 model, the barrier is a
deterministic threshold: trajectories above it are ionized, whereas trajectories
below it are neutral. In the simulation, the two populations overlap because
locations with similar smoothed density can have different ionization states
owing to their source environments, recombination histories, and the nonlocal
radiation field. The empirical barrier is therefore the equal-probability
midline of this transition region in trajectory space, rather than a sharp
physical surface \citep[see also][]{Kaurov2016,GnedinMadau2022}.

Let $\Delta_R$ denote the smoothed density trajectory measured from the simulation at smoothing radius $R$. The posterior probability that a given location at trajectory value $\Delta_R$ is neutral is
\begin{equation}
    C_{\rm neutral}(\Delta_R|R)
    =
    \frac{p_{\rm neutral}(\Delta_R|R)}
    {p_{\rm neutral}(\Delta_R|R)+p_{\rm ionized}(\Delta_R|R)} ,
\end{equation}
where $p_{\rm neutral}$ and $p_{\rm ionized}$ are the volume-weighted trajectory-value PDFs of the neutral and ionized samples. Following \citet{Kaurov2016}, the empirical barrier $B_{\rm sim}(R)$ is defined as the trajectory value at which $C_{\rm neutral}=0.5$. The width of this transition, defined as the range of $\Delta_R$ over which $C_{\rm neutral}$ moves from 0.25 to 0.75, measures how strongly the ionization state depends on variables beyond the smoothed density alone.

First and last crossings follow directly from the simulation trajectories: the first crossing is the largest smoothing radius at which a trajectory crosses $B_{\rm sim}(R)$ onto the ionized side, and the last crossing is the smallest. The endpoint of the simulation trajectory is the minimum resolved smoothing scale in the trajectory construction, not the source-halo scale, and the resolved/unresolved $R_\ell$ classification below is defined with respect to this scale.

\begin{figure*}[htb!]
    \centering
    \includegraphics[width=\textwidth]{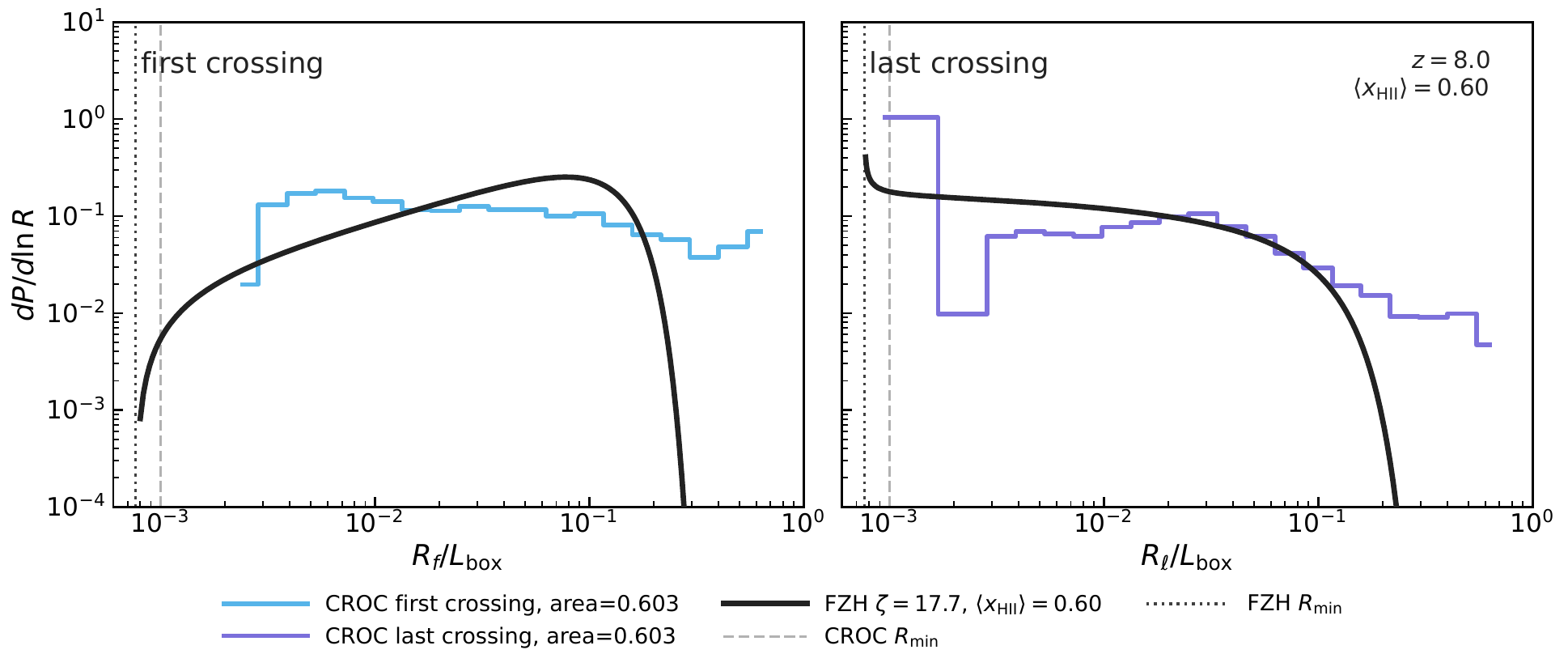}
    \caption{
    First- and last-crossing distributions measured from CROC trajectories
    and compared with the analytic FZH04 prediction at $z=8.0$ and
    $\langle x_{\rm HII}\rangle=0.60$. The left panel shows the
    first-crossing distribution and the right panel the last-crossing
    distribution. Blue is the CROC first-crossing distribution, purple is
    the CROC last-crossing distribution, and the black curves are the
    analytic FZH04 model matched to the same theoretical ionized fraction.
    Vertical dotted lines mark the minimum resolved smoothing scale in CROC
    and the source-halo endpoint in the analytic calculation. Both the
    analytic curves and the CROC histograms are normalized to the ionized
    volume fraction rather than to unity. Relative to FZH04, CROC retains
    excess probability in the large-$R_f$ and large-$R_\ell$ tails,
    consistent with nonlocal photon transport sustaining ionized regions
    whose enclosed photon budgets do not close on every scale.
    }
    \label{fig:croc_FZH04_crossing_comparison}
\end{figure*}

Having constructed the empirical barrier, I measure first and last crossings from the full CROC trajectory ensemble. Figure~\ref{fig:croc_FZH04_crossing_comparison} compares the resulting distributions with the analytic FZH04 prediction at matched $\langle x_{\rm HII}\rangle$. The CROC first-crossing distribution is broader than the analytic curve and retains substantial probability at the largest measured smoothing scales, where the analytic model has already fallen off. The last-crossing panel carries the same signature on smaller scales, with CROC developing an extended tail to larger $R_\ell$ that lies above the analytic prediction. The analytic endpoint at $S_{\min}$ and the simulation endpoint at the minimum resolved smoothing scale differ by construction, so the comparison focuses on the shape of the distributions rather than the bin heights at the smallest radii.

Photon transport explains the shape difference. In the analytic model a region is treated as ionized only when its own enclosed sources supply enough photons. In CROC, photons travel across the simulation volume, and a region need not close its own photon budget at every aperture to be ionized. Ionized regions can therefore grow to larger radii than the analytic photon-counting condition predicts, and the same nonlocal support extends to smaller scales in the last-crossing distribution. The extended resolved tail at larger $R_\ell$ indicates that a substantial fraction of the ionized volume in CROC is sustained by photons counted on apertures above the minimum resolved scale \citep{Zahn2011,GnedinMadau2022}.

\section{Resolved and Unresolved Last Crossings as Source-Counting Classes}
\label{sec:resolved_Rell_physics}

The crossing construction divides ionized regions into two classes by the scale at which the source-counting condition holds. A region with an unresolved last crossing satisfies the empirical barrier all the way down to the minimum resolved smoothing scale, meaning its local photon budget closes at the smallest aperture the analysis can probe. A region with a resolved last crossing belongs to a larger ionized bubble but falls below the empirical barrier on smaller scales, so its photon budget closes only across a larger aperture.

\begin{figure*}[htb!]
    \centering
    \includegraphics[width=\textwidth]{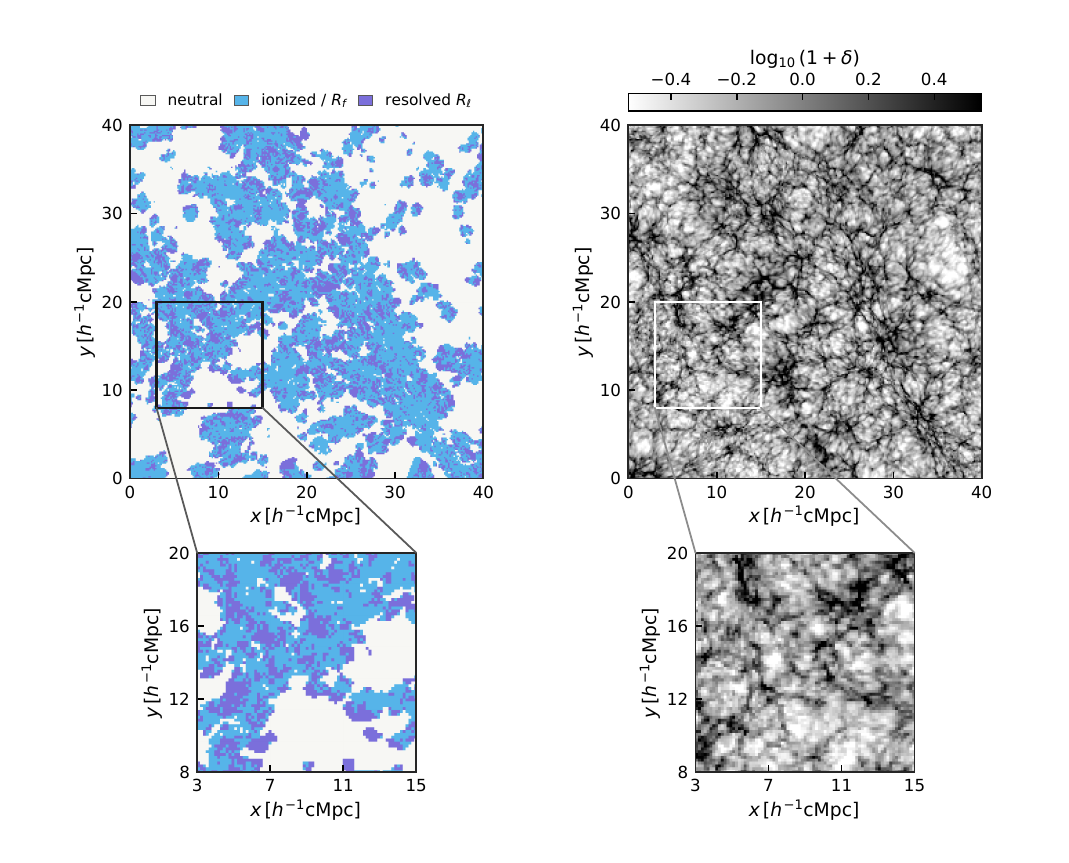}
    \caption{
    Spatial structure of the crossing classification compared with the density field.
    The left panels show a two-dimensional slice through the simulation, with neutral gas
    in white, ionized gas with an enclosing first crossing in blue, and gas with a
    resolved last crossing in purple. The right panels show the corresponding density field
    $\log_{10}(1+\delta)$ on the same slice. The lower panels zoom into the boxed region.
    Resolved-$R_\ell$ regions form coherent structures inside the ionized network, often
    surrounding remaining neutral pockets and tracing the edges of ionized regions rather
    than their interiors.
    }
    \label{fig:crossing_map_density}
\end{figure*}

Figure~\ref{fig:crossing_map_density} shows a representative slice through the simulation alongside the density field. Resolved-$R_\ell$ regions are not randomly distributed through the ionized volume: they form coherent structures that surround the remaining neutral pockets and trace the edges of ionized regions rather than their dense interiors.

\begin{figure}[htb!]
    \centering
    \includegraphics[width=\linewidth]{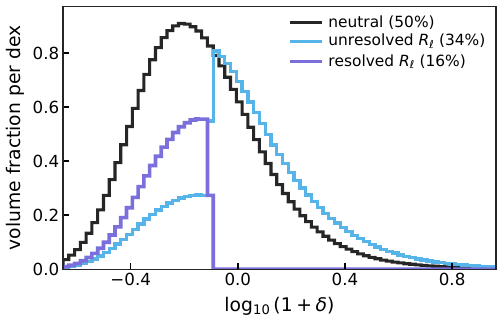}
    \caption{
    Density distributions for the three populations defined by the empirical barrier:
    neutral gas, ionized gas with unresolved last crossings, and ionized gas with
    resolved last crossings. The unresolved-$R_\ell$ population is concentrated at higher
    densities, though with a non-negligible low-density tail. The resolved-$R_\ell$ population
    is predominantly underdense.
    }
    \label{fig:density_distribution}
\end{figure}

Figure~\ref{fig:density_distribution} quantifies the density content of each class. The unresolved-$R_\ell$ population is concentrated at higher density with a non-negligible low-density tail: these regions are largely source-rich at the smallest resolved aperture and sit mainly in dense ionized structure. The resolved-$R_\ell$ population peaks well below mean density. The bulk of the underdense ionized gas in CROC therefore falls into the resolved-$R_\ell$ class: gas that is ionized but whose immediate neighborhood does not itself host enough sources to close the photon budget.

\begin{figure}[htb!]
    \centering
    \includegraphics[width=\linewidth]{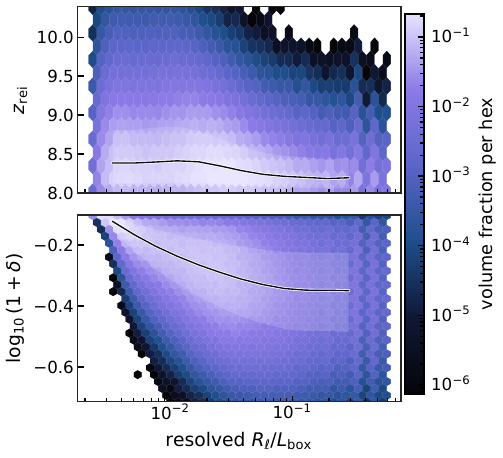}
    \caption{
    Joint distributions of resolved last-crossing scale with reionization redshift (top)
    and local density (bottom), for all resolved-$R_\ell$ regions in CROC. Hexagons show
    the binned distribution and the black curves trace the median trends. Larger
    resolved-$R_\ell$ scales correspond to lower densities and later reionization times.
    }
    \label{fig:Rell_zrei_density}
\end{figure}

Within the resolved population, $R_\ell$ varies systematically with environment. Figure~\ref{fig:Rell_zrei_density} shows the joint distributions of $R_\ell$ with local density and with reionization redshift. The median density at fixed $R_\ell$ decreases with increasing $R_\ell$, and the median reionization redshift shifts later. Regions whose source-counting condition fails on larger scales are, on average, both less dense and reionized at later times.

\begin{figure}[htb!]
    \centering
    \includegraphics[width=\linewidth]{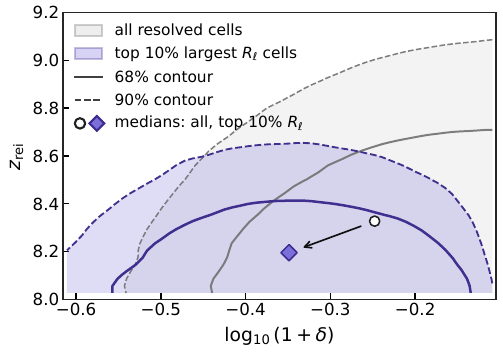}
    \caption{
    Joint distribution of local density and reionization redshift for all resolved-$R_\ell$
    regions and for the top 10\% largest-$R_\ell$ subset. Contours enclose 68\% and 90\%
    of each population, and markers show the medians. The top 10\% largest-$R_\ell$
    regions shift coherently toward lower density and lower $z_{\rm rei}$.
    }
    \label{fig:joint_density_zrei}
\end{figure}

Figure~\ref{fig:joint_density_zrei} shows where the largest-$R_\ell$ regions
sit in the joint density--reionization-time plane. The top 10\% largest-$R_\ell$
subset shifts coherently to lower density \emph{and} later reionization
relative to the full resolved sample, occupying a distinct corner of the joint
plane rather than the tail of either variable alone. In the inside-out
picture, source-rich overdense regions ionize first, and ionizing photons
subsequently reach the lower-density gas between them; the largest-$R_\ell$
regions preferentially select this later-ionized, low-density gas.

Unresolved-$R_\ell$ regions, by contrast, are concentrated at higher density
and are source-rich at the smallest resolved scale: their enclosed
photon-counting condition remains satisfied on every scale probed. The bulk
of the underdense ionized gas in CROC instead belongs to the
resolved-$R_\ell$ class, consistent with ionization by sources distributed
through the denser regions surrounding it.

\section{Observational Implications}
\label{sec:observational_program}

\begin{figure*}[htb!]
\centering
\includegraphics[width=\textwidth]{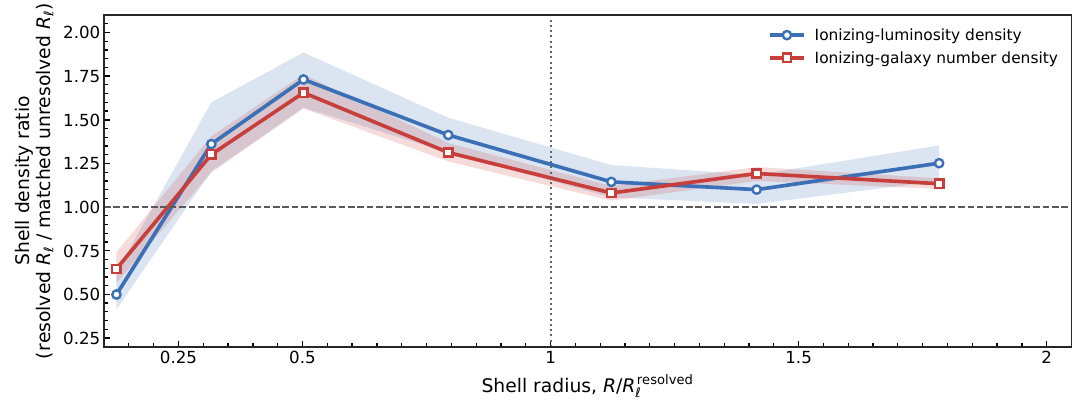}
\caption{
Radial ionizing-source environments of resolved-$R_\ell$ regions at $z=8.0$,
relative to matched ionized unresolved-$R_\ell$ regions. Blue circles show
the ratio of mean ionizing-luminosity density, and red squares show the
ratio of mean ionizing-galaxy number density, in radial shells. Each
unresolved-$R_\ell$ region is evaluated using the $R_\ell$ of its matched
resolved-$R_\ell$ partner. The vertical dotted line marks $R=R_\ell$, and
the horizontal dashed line marks equal source densities. Shaded regions
show $16$th--$84$th percentile intervals from paired bootstrap resampling. Resolved-$R_\ell$ regions are deficient in nearby sources but surrounded by an
excess at intermediate radii relative to matched unresolved regions, indicating that their ionization is sustained by
a more extended source environment.}
\label{fig:Rell_source_environment}
\end{figure*}

Figure~\ref{fig:Rell_source_environment} shows that resolved-$R_\ell$ regions
have a distinct source environment. I match each resolved-$R_\ell$ region to an
ionized region without a distinct resolved last crossing, holding fixed the
local density, the enclosing first-crossing scale $R_f$, and the reionization
redshift. In the resolved-$R_\ell$ sample, the innermost shell,
$R<0.25R_\ell$, has an ionizing-luminosity density lower by approximately a
factor of two, and a galaxy number density lower by roughly one third. Both
ratios rise above unity outside the central shell and peak near
$R\simeq0.5R_\ell$, where the luminosity and number densities are enhanced by
roughly 75\% and 65\%, respectively.

The classification therefore retains information about the source distribution
beyond what $\delta$, $R_f$, and $z_{\rm rei}$ already contain. A resolved
last crossing marks a trajectory that satisfies the source-counting condition
down to $R_\ell$ but not on smaller resolved scales. The measured profile is
consistent with this interpretation: resolved-$R_\ell$ regions have fewer
sources near their centers and more sources at intermediate radii than matched
unresolved regions. The last-crossing classification therefore separates
source environments that $\delta$, $R_f$, and $z_{\rm rei}$ alone do not
distinguish.

Resolved-$R_\ell$ regions are also preferentially underdense, linking this
source geometry to Ly$\alpha$ transmission. \citet{Zhu2024} found that low gas
density is the primary condition producing transmission spikes in CROC, while
the ionizing radiation field plays a secondary role and nearby galaxies add
little predictive information once the gas state is known. These results
motivate the hypothesis that resolved-$R_\ell$ regions preferentially host
transmission spikes. Their low central density reduces the Ly$\alpha$ opacity,
while their embedding within a larger source-supported ionized region allows
the gas to remain highly ionized. A transmission spike may therefore select a
locally source-poor region whose ionization is sustained by its larger-scale
environment, rather than a location centered on an unusually rich
concentration of galaxies.

This picture can be tested with transmission-selected galaxy--IGM
measurements, although transmission selection does not map uniquely onto
$R_\ell$. \citet{YongdaZhu2026} provide a current example: the regions of peak
transmission at $z\simeq5.7$ along both QSO sightlines in their study are
underdense in [O~{\sc iii}] emitters, with galaxy densities less than half the
cosmic mean, whereas another transmissive interval lies in an environment of
at least average galaxy density. Their two-dimensional IGM--galaxy
cross-correlation shows a tentative off-axis enhancement of transmission near
$(\Delta r,\Delta d)\simeq(0.8,0.6)\lambda_{\rm mfp}$. High transmission can
therefore occur away from locally galaxy-rich environments, while the
galaxy--IGM relation retains structure at finite separation.

The corresponding observational test is the galaxy density around
transmission-selected locations. If these locations preferentially sample the
resolved-$R_\ell$ population, their stacked galaxy profile should show a
central deficit and an excess at finite separation relative to matched control
locations along the Ly$\alpha$ forest. Mock Ly$\alpha$ spectra and galaxy
catalogs are required to test this association directly and to map the
three-dimensional profile in Figure~\ref{fig:Rell_source_environment} into the
projected, anisotropic geometry of the observations, accounting for redshift
errors, luminosity limits, survey masks, and the selection of transmissive
regions.

\section{Summary}
\label{sec:summary}

The FZH04 photon-counting barrier contains complementary first- and
last-crossing scales. The first crossing identifies the largest scale on which
the enclosed sources can ionize the enclosed gas, while the last crossing
identifies the smallest scale on which that photon budget still closes.
Trajectories that remain above the barrier to the endpoint have no resolved
last crossing and satisfy the photon-counting condition on every scale probed.
An interior last crossing at $R_\ell$ marks gas that is externally ionized on
smaller resolved scales: its local source population is insufficient, while
sources enclosed by the larger surrounding volume close the photon budget.

In the analytic model, the large-scale extent of the last-crossing
distribution moves outward with the characteristic first-crossing scale while
remaining weighted toward smaller radii, producing approximately self-similar
evolution. The empirical CROC barrier produces a broader first-crossing
distribution and a more extended tail toward large $R_\ell$ than the analytic
FZH04 model. This difference reflects nonlocal photon transport: ionized gas
in the simulation need not close its photon budget within every aperture
because photons propagate from sources on larger scales. Resolved-$R_\ell$
regions form coherent structures around remaining neutral pockets and along
the edges of ionized regions. They are predominantly underdense, and within
the resolved population, larger $R_\ell$ selects lower density and later
reionization.

The matched source-environment analysis shows that $R_\ell$ retains
information beyond local density, enclosing first-crossing scale, and
reionization redshift. At fixed values of these variables, resolved-$R_\ell$
regions have approximately half the ionizing-luminosity density and a galaxy
number density lower by about one third in their central shells than matched
unresolved regions. At intermediate radii, both source densities exceed those
of the matched sample, with the luminosity-density enhancement reaching
approximately 75\%. The last-crossing classification therefore identifies a
source geometry with a locally source-poor center embedded within a richer
surrounding environment.

The results motivate a physical picture for Ly$\alpha$ transmission spikes.
Low gas density is the primary condition producing spikes in CROC, and resolved-$R_\ell$ regions combine low density with a
deficit of nearby sources and an excess at larger radius. Transmission spikes
may therefore preferentially arise in these externally ionized regions: low
density reduces the Ly$\alpha$ opacity, while sources distributed over larger
scales maintain the ionized state. This picture predicts a central galaxy
deficit and an excess at finite separation around transmission-selected IGM
locations. The galaxy underdensities and tentative finite-separation
transmission signal reported by \citet{YongdaZhu2026} bear on this prediction.

\appendix

\section{Monte Carlo Validation}
\label{app:mc_validation}

The Monte Carlo calculation provides an independent check on the deterministic last-crossing solver by generating sharp-$k$ trajectories directly, recording their first and last crossings, and comparing the resulting histogram with the analytic prediction.

Sharp-$k$ trajectories are generated on a grid of variances,
\begin{equation}
    0=S_0<S_1<\cdots<S_N=S_i ,
\end{equation}
starting from $\delta(S_0)=0$ and incrementing by
\begin{equation}
    \delta(S_{j+1})-\delta(S_j)
    =
    \sqrt{S_{j+1}-S_j}\,G_j ,
\end{equation}
where $G_j$ is a standard normal random number. Each interval of variance contributes an independent Gaussian increment, reproducing the sharp-$k$ random walk.

The FZH04 barrier $B_j=B(S_j,z)$ is evaluated at the same grid points, and a grid point is marked as satisfying the FZH04 condition when $\delta(S_j)\ge B_j$. The smallest such $S_j$ gives a grid-level estimate of the first crossing and the largest gives a grid-level estimate of the last crossing.

A finite grid misses crossings between grid points; this discretization error is corrected with a Brownian bridge. Within a single interval from $S_j$ to $S_{j+1}$, the signed distances below the barrier at the endpoints are
\begin{equation}
    y_j=B_j-\delta_j,
    \qquad
    y_{j+1}=B_{j+1}-\delta_{j+1}.
\end{equation}
If either endpoint has $y<0$, the trajectory is already above the barrier at a grid point. If both have $y>0$, the trajectory is below the barrier at both grid points but may still cross between them. Approximating the barrier as linear across the interval, the bridge crossing probability is
\begin{equation}
    P_{\rm bridge}
    =
    \exp\!\left[
    -\frac{2y_jy_{j+1}}{S_{j+1}-S_j}
    \right].
\end{equation}
A uniform random number $U\in[0,1]$ is drawn for each such interval, and the interval is treated as containing a crossing when $U<P_{\rm bridge}$. The only approximation is replacing the true barrier by a straight line across one step.

The first crossing is the earliest grid point or bridge interval where a crossing is recorded, and the last crossing is the latest. Trajectories that remain above the barrier at $S_i$ are assigned to the endpoint contribution $S_\ell=S_i$; otherwise the last crossing feeds the interior histogram. With $N_{\rm traj}$ trajectories and $N_n$ trajectories whose interior last crossing falls in radius bin $n$, the density is
\begin{equation}
    \frac{dP}{d\ln R}
    \simeq
    \frac{N_n/N_{\rm traj}}
    {|\ln(R_{n,{\rm right}}/R_{n,{\rm left}})|}.
\end{equation}
The endpoint probability is measured separately as the fraction of trajectories with $\delta(S_i)\ge B_i$.

\begin{figure}
    \centering
    \includegraphics[width=\linewidth]{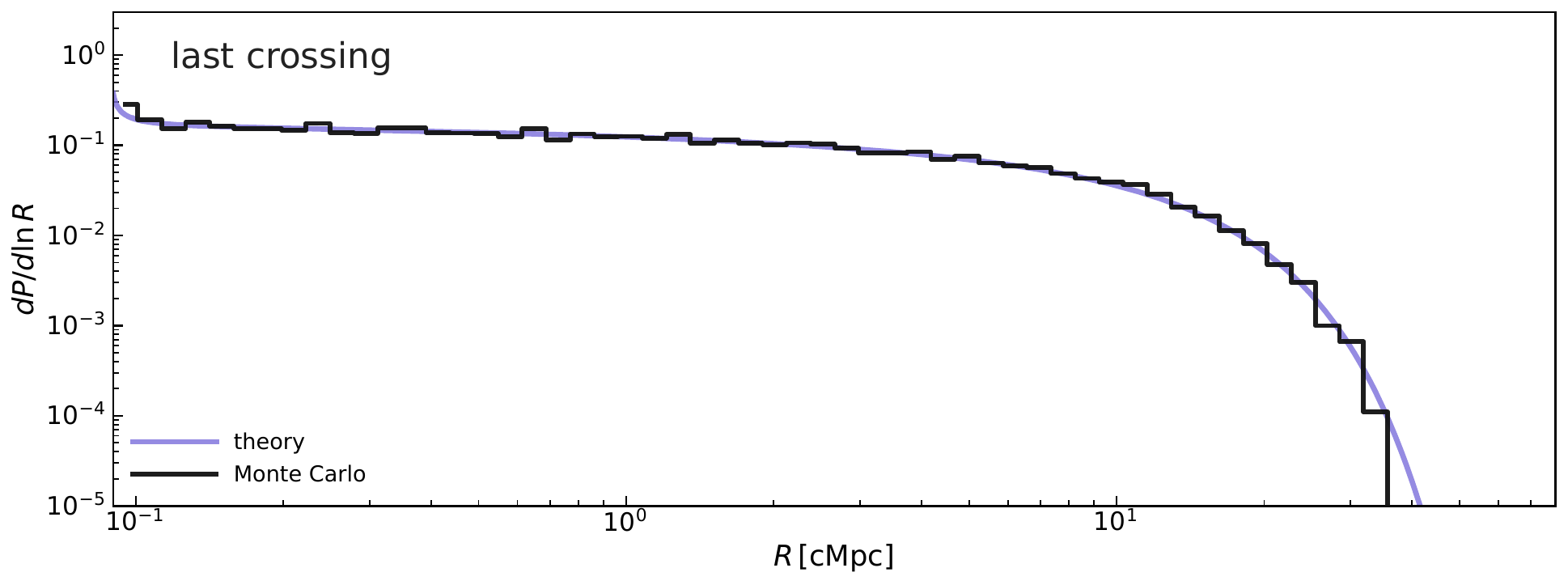}
    \caption{
    Monte Carlo validation of the finite-endpoint last-crossing calculation.
    The deterministic solution agrees with direct sharp-$k$ random-walk realizations,
    confirming that the recursive solver reproduces the last-crossing statistics.
    }
    \label{fig:mc_validation}
\end{figure}

Figure~\ref{fig:mc_validation} compares the deterministic solution against direct Monte Carlo trajectories. The two methods agree, confirming that the binned finite-endpoint solver reproduces the crossing statistics of the sharp-$k$ process. The deterministic solver is smoother and more efficient, especially in the tails; the Monte Carlo follows the trajectories explicitly and provides an independent cross-check of the crossing definitions.

\begin{acknowledgments}
I thank Nick Gnedin for helpful discussions that significantly improved this paper. I also thank my thesis committee members---Wayne Hu, Alex Ji, Irina Zhuravleva, and Ellen Zweibel---for their valuable comments and discussions.
\end{acknowledgments}

\bibliography{main}{}
\bibliographystyle{aasjournal}

\end{CJK*}
\end{document}